\documentclass[]{article}

\usepackage{epsfig}

\newcommand{\AmS}{{\protect\the\textfont2
  A\kern-.1667em\lower.5ex\hbox{M}\kern-.125emS}}

\begin{document}

\title{Analytic calculation of the mass gap in
U$(1)_{2+1}$ lattice gauge theory.}

\author{John A.L. McIntosh and Lloyd C.L. Hollenberg.\\ 
        \\ Research Centre for High Energy Physics,\\
	School of Physics, \\
	University of Melbourne,\\
	Victoria 3010 Australia.}

\maketitle

\begin{abstract}
An analytic calculation of
the photon mass gap $M$ of compact U$(1)_{2+1}$
in the Hamiltonian formalism is performed utilizing the first four
Hamiltonian moments with respect to a one-plaquette mean
field state in the plaquette expansion method. Scaling of $M$ is clearly evident at and
beyond the transition from strong to weak coupling. The scaling
behaviour agrees well with the range of results from numerical calculations.
\end{abstract}

\section{Introduction}
In QCD, the most fundamental
quantities one wishes to calculate are the states of the mass
spectrum of the theory. In lattice gauge theory, such
non-perturbative calculations are carried out numerically in the
path integral representation to some level of approximation by
evaluating the correlation functions by Monte-Carlo simulation.
Typically, the computational resources employed in this approach
are enormous as one must work with large lattices and spacings which
are fine enough to detect the continuum behaviour of the theory in
question. Hence, non-perturbative analytic results are of considerable interest.
In this context, the Hamiltonian formalism is appealing.
One often works
directly on infinite lattices, fermions are simpler to incorporate
(no fermion determinant) and 3D as opposed to 4D lattices (time is
continuous). 
The lack of systematic
and applicable methodology has prevented these features from being
exploited or investigated - hence, the importance of new
methodologies.
Here we use a novel
approach -- the plaquette expansion -- based on a large volume
expansion of Lanczos tridiagonalization.

\section{Compact U(1) in 2+1 Dimensions}
Compact U$(1)_{2+1}$ has become the testing ground for
various lattice Hamiltonian procedures
as it exhibits scaling behaviour
similar to the more complex and physically interesting non-abelian
lattice gauge theories in $3+1$ dimensions.
The Kogut-Susskind Hamiltonian is~\cite{kogsuss75}:
\begin{equation}
H=\frac{g^2}2\sum_{l}{\hat E}^2_{l}
 + \frac1{2g^2}\sum_{p}\left[2-({\hat U}_{p} + {\hat U}_{p}^{\dag})\right]\ ,
\label{eq:hamu1}
\end{equation}
where $g$ is the dimensionless coupling constant.
The strong-coupling limit is defined by~$g\rightarrow\infty$ and
the weak-coupling limit by $g\rightarrow 0$.
The electric field operator, ${\hat E}_{l}$, and link operator, ${\hat U}_{l}$
obey the commutator relations
$[{\hat E}_{l},{\hat U}_{l}]={\hat U}_{l}$ and
$[{\hat E}_{l},{\hat U}_{l}^{\dag}]=-{\hat U}_{l}^{\dag}$ and
the plaquette~operator ${\hat U}_{p}$ acts on the links around the
smallest closed (Wilson) loop or square on the lattice
${\hat U}_{p}={\hat U}_{1}{\hat U}_{2}{\hat U}_{3}^{\dag}{\hat U}_{4}^{\dag}$.

One of the
quantities of interest in this model is the anti-symmetric or
photon mass gap, $M$~(~$\equiv M_{0^{+-}}$~), which is given by
the difference in energies between the lowest state in
the~$0^{+-}$ sector and the vacuum. The scaling behaviour
for the mass gap is expected to be~\cite{gopmack82}:
$M^{2} a^{2}\equiv\beta \mbox{exp}\{-k_{0}\beta\ +\ k_{1}\}$,
where $k_{0}$ and $k_{1}$ are constants, $a$ is the lattice spacing and
$\beta =1/g^{2}$.
The scaling parameters $k_{0}$ and $k_{1}$ are not known exactly but
have been computed by various numerical techniques: $k_{0}\sim 4.1 - 6.345$
and $k_{1}\sim 4.369 - 6.27$ (See~\cite{toPRL} for summary).

\section{Plaquette Expansion}

Beginning with a trial state $|\psi_{1}\rangle$
which has the desired symmetries of the state
of interest, the Lanczos recurrence generates a basis
\begin{eqnarray*}
|\psi_{n}\rangle=\frac{1}{\beta_{n-1}}
\left[\left(H-\alpha_{n-1}\right)|\psi_{n-1}\rangle
-\beta_{n-2}|\psi_{n-2}\rangle\right],
\end{eqnarray*}
where $\alpha_{n}=\langle\psi_{n}|H|\psi_{n}\rangle$ and
$\beta_{n}=\langle\psi_{n+1}|H|\psi_{n}\rangle$ are the matrix
elements of the Hamiltonian in tri-diagonal form.

The matrix elements are able to be written in terms of Hamiltonian
moments $\langle H^{n}\rangle\equiv\langle\psi_{1}|H^{n}|\psi_{1}\rangle$.
The connected part of the Hamiltonian moment is proportional to the
volume of the system $\langle H^{n}\rangle_{c}=e_{n}\,N$.
Therefore, one may re-express the matrix elements in terms of the
connected vacuum coefficients $e_{n}$.\ Extensivity of the problem
leads to the following plaquette expansions in $1/N$~\cite{holl93}:
\begin{equation}
{\alpha_n\over N}=e_{1} + s\left[\frac{e_{3}}{e_{2}}\right] +
O(s^2)\equiv {\bar \alpha}(s), \label{eq:avac}
\end{equation}
\begin{equation}
{\beta_n^{2}\over N^2}=s e_{2} + \frac12
s^{2}\left[\frac{e_{2}e_{4}-e^2_{3}}{e^2_{2}} \right] +
O(s^3)\equiv {\bar \beta^2(s)} \label{eq:bvac}
\end{equation}
where $s\equiv n/N$. In the bulk limit ($n,N\rightarrow\infty$),
keeping $s$ fixed, one may perform  the exact diagonalization of
the Lanczos tri-diagonal matrix for the ground state energy
density $\varepsilon_{0}$ analytically~\cite{hwaf95}:
$\varepsilon_{0}= {\rm inf}\, ({\bar \alpha}(s) - 2{\bar\beta(s)})\,$.
For example, for 4th order moments we have:
\begin{equation}
\varepsilon_{0}[4] = e_{1} + \frac{e_{2}^{2}}{e_{2} e_4 -
e_{3}^{2}}\,\left[ \sqrt{3 e_{3}^{2}-2 e_{2} e_{4}}-e_{3}\right].
\label{eq:E0}
\end{equation}
For excited states the Hamiltonian moments have the form
$\langle H^{n}\rangle^{(S)}_{c}=e_{n} N + m_{n}^{(S)}$.
Thus, in a similar fashion to $\varepsilon_{0}$, one
can derive expressions for approximants, $M^{(S)}_{0}[r]$, to the
mass gap. Again using up to 4th order moments we have~\cite{hollwilwit95}:
\begin{equation}
M^{(S)}_{0}[4]=\sum^{4}_{n=1}\,m^{(s)}_{n}\,e_2^{n-1}\,F_{n}(e_{2},e_{3},e_{4}),
\label{eq:mfmg}
\end{equation}
where the vacuum moment functions, $F_{n}$, are simple algebraic
forms given in~\cite{hollwilwit95}. 

\section{One Plaquette Mean-Field State}
Initially, lattice Hamiltonian calculations
employed the simplest gauge invariant trial state --- the
strong-coupling vacuum $|0\rangle$ (defined as the state
satisfying ${\hat E}_{l}|0\rangle =0$ for every link). This is the
perturbative starting point for series calculations, which are
then extrapolated to weak coupling. Although the calculation of
moments with respect to this state can be carried out to
relatively high order, typically, one finds that this state is
simply inadequate to explore the non-perturbative weak-coupling
regime of the theory. In the case of the plaquette expansion, a
window of scaling in terms of the expansion order was evident, but
higher order results became problematic~\cite{macholl97}.

\begin{figure}[htbp]
\hbox{\hspace{20mm}\parindent=0mm
        \vtop{%
                \psfig{file=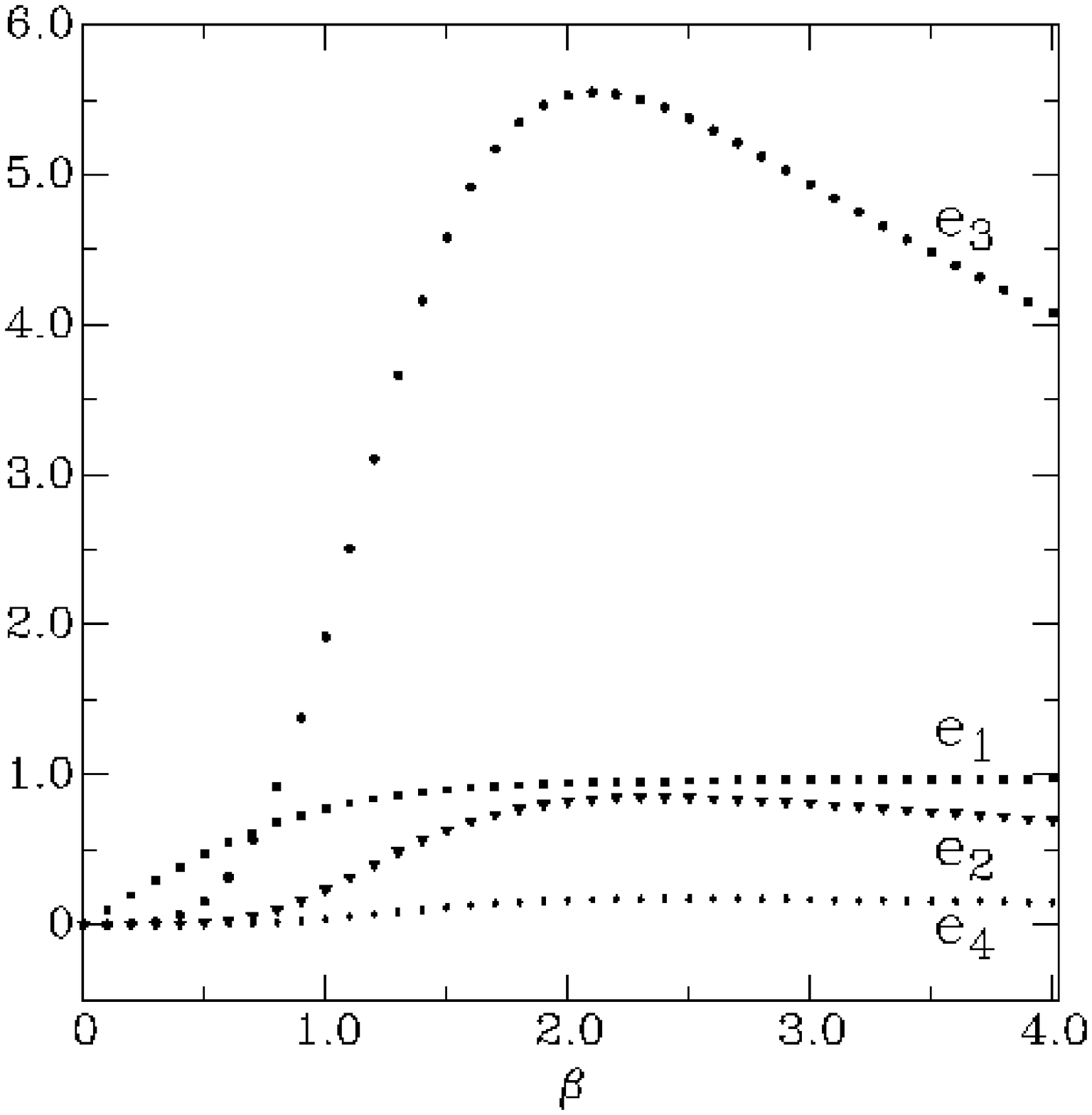,width=80mm}
                \vbox{\hsize=80mm}
        }
}
\caption{Ground state moments, $e_{1}\rightarrow e_{4}$.}
\label{fig:gsmom}
\end{figure}

Here we utilized a one plaquette exponential
trial state, that has the correct strong and weak coupling behaviour for the
vacuum energy density -- the so-called {\it mean-field} state~\cite{bish96} --
defined in the so-called $U$ representation as:
\begin{equation}
\psi_1[U_p] = \exp\left\{\frac{\lambda}{2}
\sum_{p}\left\{{U}_{p}+{U}_{p}^{\dagger}\right\}\right\}
\end{equation}
where $\lambda$ is determined variationally from $\langle H\rangle$.
As the state is constructed from plaquette variables, $U_{p}$, it
is automatically gauge invariant. One expects the true ground
state of the system to have contributions from all possible Wilson
loops -- an important feature of the plaquette expansion method is
that, by its very nature, larger sized Wilson loops (larger
clusters) are systematically introduced.

\begin{figure}[htbp]
\hbox{\hspace{20mm}\parindent=0mm
        \vtop{%
                \psfig{file=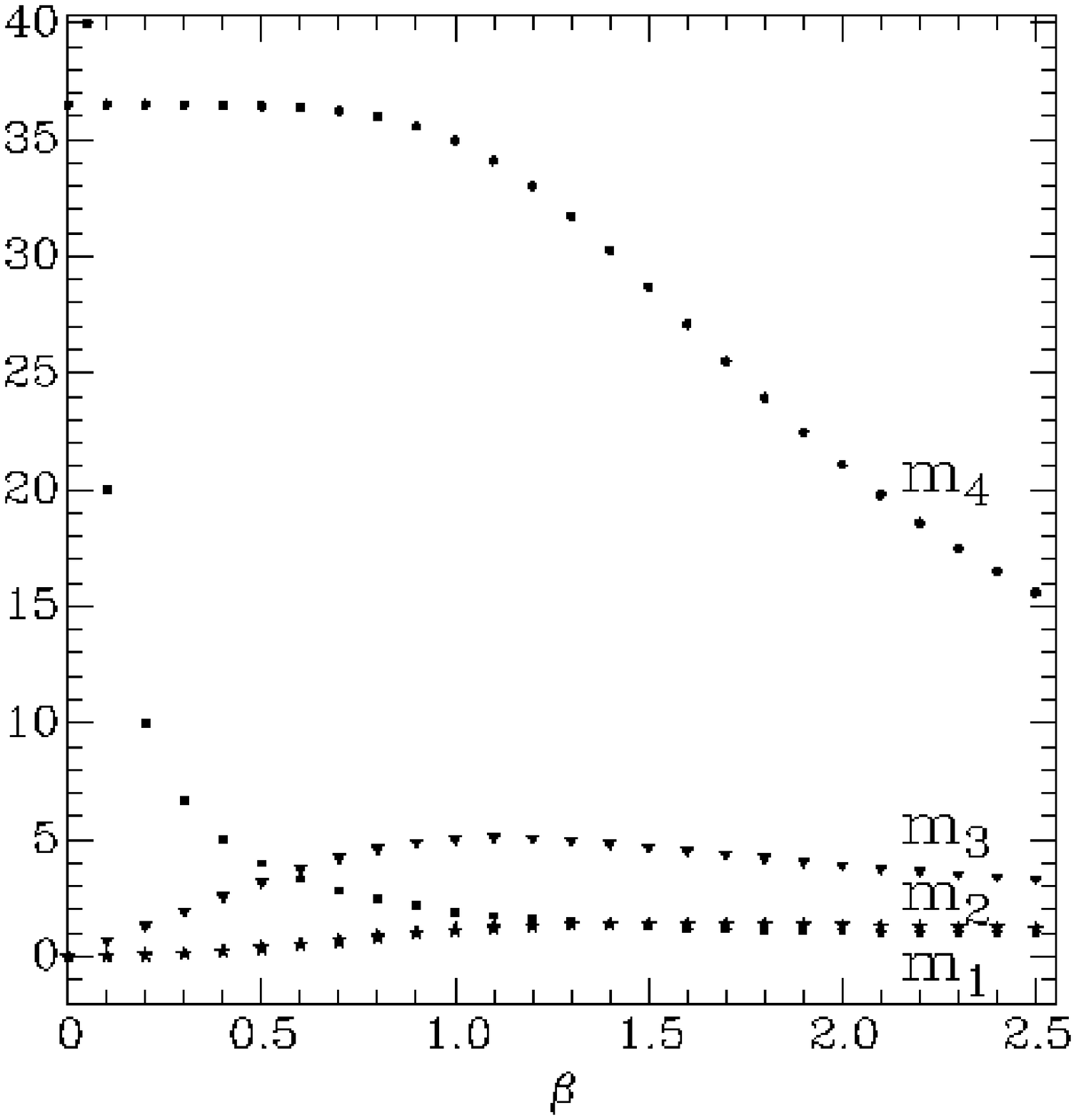,width=80mm}
                \vbox{\hsize=80mm}
        }
}
\caption{Anti-symmetric moments, $m_{1}\rightarrow m_{4}$.}
\label{fig:asmom}
\end{figure}

Using a diagrammatic method (details to be presented elsewhere)
we were able to derive the first four
one-plaquette connected moments for both the ground
state and anti-symmetric state. The expressions become rather
long, so for brevity these are illustrated in Figures \ref{fig:gsmom} and
\ref{fig:asmom}, 
along the curve $\lambda(\beta)$. These moments
immediately give analytic approximates for vacuum energy and
the mass gap.

\section{Results and Discussion}
\begin{figure}[htbp]
\hbox{\hspace{10mm}\parindent=0mm
	\vtop{%
               	 \psfig{file=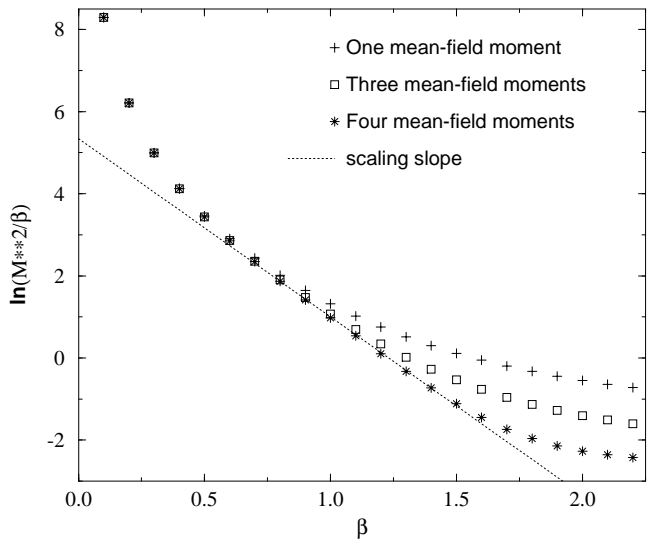,width=100mm}
		\vbox{\hsize=110mm}
	}
}
\caption{Mass gap results applying one, three and four one-plaquette
Hamiltonian moments. Scaling is evident for the four order moments curve
from $\beta=0.7$ to $1.5$.}
\label{fig:compmfmg}
\end{figure}

The plaquette expansion results are presented 
in Figure~\ref{fig:compmfmg}. Clearly scaling is evident for
inverse-coupling values $\beta=0.7$ to $1.5$, passing the
transition point at $\beta=1.0$. The scaling form we find is given
by $M^{2}=\beta\,\mbox{exp}\left(-4.34\beta+5.34\right)$, and
agrees well with other estimates~(summarized in~\cite{toPRL}).

Although we have presented
results for only the first four moments (for which the calculation
is entirely analytic) it should in principle be possible to derive
higher moments with respect to this state. Indeed we have
preliminary results for the ground state energy density to sixth
order giving high accuracy over a large range of couplings. 

For the future application of this method to $\mbox{QCD}_{3+1}$,
the main obstacle to
overcome is the calculation of the moments in 3+1 dimensions. The
appeal of 2+1 systems as toy models is that the transformation
from link variables to plaquette variables is trivial and makes
the integrations tractable -- hence the analytic work reported here.
In 3+1 dimensions it is well known that the
transformation cannot be carried out in closed form due to the
appearance of Bianchi identities. However, the calculation of
moments in 3+1 dimensions by Monte-Carlo is actually a relatively
small scale numerical exercise. Because we are dealing with a
cluster expansion (we need only connected moments) the lattices
required are fairly modest in extent and, by definition, only 3
dimensional. The integrands become complicated as larger
correlations are included, which means that the statistics must be
very good, however, preliminary calculations have demonstrated
that the required precision is possible to achieve without
supercomputing resources.


\end{document}